\documentclass[usegraphicx,useAMS]{mn2e}

\newcommand{\bgr}{\bibitem[\protect\citename{dummy }1893]{dum}}

\newcommand{\etalc}{et~al.}

\newcommand{\mum}{\,$\umu$m}

\begin{document}

\title[Submillimetre observations of blazars]{Observations of flat-spectrum
  radio sources at $\lambda 850\umu$m from the James Clerk Maxwell Telescope I. April 1997--April 2000}


\author[E. I. Robson et al.]{E. I. Robson,$^{1,2}$ J. A. Stevens$^3$ and
  T. Jenness$^1$ \\
$^1$ Joint Astronomy Centre, 660 N. A`oh\={o}k\={u} Place, University Park,
Hilo, Hawaii, 96720, USA \\
$^2$ Centre for Astrophysics, University of Central Lancashire, Preston, PR1
2HE \\
$^3$ Mullard Space Science Laboratory, University College London, Holmbury
St. Mary, Dorking, Surrey, RH5 6NT \\
}

\date{Accepted 2001 June 12. Received 2001 March 13}

\maketitle

\begin{abstract}
  
  Calibrated data for 65 flat-spectrum extragalactic radio sources are
  presented at a wavelength of 850\mum\ covering a three-year period from
  April 1997. The data, obtained from the James Clerk Maxwell Telescope using
  the SCUBA camera in pointing mode, were analysed using an automated pipeline
  process based on the Observatory Reduction and Acquisition Control - Data
  Reduction (ORAC-DR) system. This paper describes the techniques used to
  analyse and calibrate the data and presents the database of results along
  with a representative sample of the better-sampled lightcurves.

\end{abstract}


\section{Introduction}

The James Clerk Maxwell Telescope (JCMT) is a 15-m diameter radio telescope
situated on the 4.2~km peak of Mauna Kea in Hawaii. The facility is equipped
with a suite of common-user instruments, including the Submillimetre
Common-User Bolometer Array, SCUBA (Holland et al. 1999). The data reported in
this paper comprise 2238 flux measurements from pointing observations of 65
flat-spectrum extragalactic radio sources. The observations have been reduced
using the automatic SCUBA data reduction pipeline (Jenness et al. 2001)
encompassing on-line atmospheric extinction correction using quasi-continuous
monitoring data from the Caltech Submillimetre Observatory (CSO). Lightcurves
for the most frequently sampled sources are presented.

Astronomical programmes aimed at understanding the emission mechanisms in this
class of source are inevitably data limited, especially for time-dependent
multi-wavelength variability studies. The millimetre-submillimetre region is
very important for these, and so these data, like those at a longer wavelength
from IRAM (Steppe et al. 1988, 1992, 1993; Reuter et al. 1997) will be a most
valuable database for reference and use. Furthermore, a significant number of
these sources have no previously published observations at submillimetre
wavelengths.

\section{The sources}

Flat-spectrum radio sources are AGNs with relativistic jets. Monitoring of the
emission gives insights into the jet emission, and can be particularly
revealing during flaring behaviour. Although the sources fall into categories
labelled by classes such as radio-loud quasars and BL Lacertae objects, for
the purpose of this paper these sub-classifications are not important. Rather,
this paper is a catalogue of observations of a heterogeneous sample of
flat-spectrum extragalactic radio sources covering a three-year period; their
commonality being that they are all relatively bright at 850\mum\ and that
they are visible from the JCMT.

The flat-spectrum radio sources have synchrotron spectra that extend into the
submillimetre with an index, $\alpha$, that is typically between 0.5--0.7
($S\nu\propto\nu^{-\alpha}$) at submillimetre wavelengths (e.g. Gear et al.
1994). The flat nature of the radio spectrum is believed to be due to the
superposition of a number of self-absorbed synchrotron emitting blobs within
the jet that emanates from the central engine. This view is supported by very
high spatial resolution radio studies that have been able to deconvolve a
number of these components (e.g. Marscher 1988).  The highest frequency
synchrotron component, which often occurs in the submillimetre, comes from the
innermost part of the jet according to most models (e.g. Marscher \& Gear
1985). Extensive studies in the millimetre/submillimetre regime have
concentrated on monitoring the detailed flux variations and spectral evolution
of this component (e.g. Robson et al. 1983; Valtaoja et al 1998; Stevens et
al. 1996, 1998)

Another important aspect of the synchrotron emission in these sources is the
ability of the relativistic electrons to interact with the local photon field
to produce higher frequency radiation, such as X- and $\gamma$-rays, through
the inverse-Compton process (e.g. Sikora, Begelman \& Rees 1994).  Progress in
determining the precise mechanism of high-energy photon production is
currently only achievable through temporal correlations of the emission
between what is believed to be the low-energy seed photons and the high-energy
progeny. One of the major problems in this research, however, is the scarcity
of data at all wavelengths. Although some coordinated campaigns involving
satellites have been undertaken (e.g. McHardy et al. 1999), nevertheless, the
lack of data is a crucial element in stifling progress. The data in this paper
will go some way to assisting these programmes in terms of the submillimetre
flux record.

Although astronomical programmes on the JCMT have been undertaken with the
express purpose of monitoring these variable extragalactic sources, especially
blazars, the data contained in this paper are not of this kind.  Blazar
monitoring was routinely carried out using the previous single-pixel
photometer, UKT14 (Stevens et al. 1994) but has declined with the advent of
SCUBA. For the monitoring programmes the source samples were carefully
selected, however, the sources included here make up a very ad hoc sample and
the sampling rate varies significantly. Some sources are observed much more
frequently because they are either very bright and useful for set-up and
monitoring observations of pointing and focusing (e.g. 3C~279), they are
well-placed for large observing programmes (e.g. 1308+326 for the Hubble Deep
Field North observations - Hughes et al. 1998), or they are one of the few
pointing sources in that part of the sky.

\section{The observations}

Although the rms absolute pointing of the JCMT is typically 1.5 arcsec, it is
almost universal practice to use a local pointing source for accurate source
registration. Pointing sources should be bright and compact with respect to
the 850\mum\ beam ($\sim$14.5 arcsecs). Because there are few planets or
secondary calibration sources (see Sandell 1994) available, most pointing
sources are bright, flat-spectrum, radio sources. These give excellent sky
coverage and allow pointing observations to be undertaken in times of order 1
minute. The observations in this paper are entirely from these pointing
observations using the continuum camera, SCUBA, and at a single wavelength of
850\mum. The observation technique is described below.

Because the SCUBA pixels are fed by feedhorns with spacing of 2F$\lambda$,
where F is the focal ratio and $\lambda$ is the wavelength of observation, the
array does not instantaneously fully sample the image plane.  In order to
obtain a fully sampled image and hence a `picture', the image is moved across
the array in a precisely determined pattern. This is achieved using the
telescope secondary, which is rapidly moved through a jiggle pattern. For a
single wavelength, such as 850\mum\ considered here, this is a 16 position
jiggle, each jiggle position offset by $\sim$6 arcsec. A complete jiggle map
includes the telescope nod position to remove imbalances and is made up of
2$\times$16 jiggles and takes 32 seconds to complete. Successive jiggle maps
are added together to build-up the required signal-to-noise for the final
image, a 2.3 arcmin diameter picture of the sky.

Pointing is a 16 position jiggle-map where the source is fitted with a
centroid to determine its precise location and hence the telescope local
offsets.  Although a default pointing observation consists of two complete
jiggle-maps, the brightest sources, like 3C~279, are often observed with one;
conversely as many as ten maps have been used for the faintest sources.
Taking the measured range of noise equivalent flux density (NEFD) measured for
SCUBA's 850\mum\ filter ($\sim75-110$~mJy\,Hz$^{-1/2}$) the observed rms noise
level on the maps is thus expected to vary between about 20 and 50 mJy.
Therefore, providing they can be well calibrated, pointing observations
represent a valuable archive of data for the variable extragalactic sources.

All observations from the JCMT are archived, both on-site and at the Canadian
Astronomical Data Centre (CADC).  The next section discusses the processing of
the pointing data from this archive to determine the fluxes for the
extragalactic radio sources. The paper by Jenness et al. (2001) discusses the
general case of the pipeline process of data extraction and reduction in much
greater detail.

\section{The data reduction technique}

\subsection{Introduction}

Converting the pointing observations into a useful source of data requires
three critical steps. The first is to extract the data in such a way that they
are not affected by local parameters (such as telescope focus condition). The
second is to correct for the atmospheric extinction, while the third is the
flux density calibration.

Because of the large quantity of data, it was essential that an automated
technique was used. All data were thus reduced using the ORAC-DR data
reduction pipeline developed for SCUBA and the JCMT (Jenness et al. 2001).
This procedure, along with the data analysis techniques are described below.

\subsection{The pipeline: data processing and calibration}

ORAC is the Observatory Reduction and Acquisition Control system developed for
the United Kingdom Infrared Telescope (UKIRT). ORAC-DR (Economou et al. 1999;
Jenness \& Economou 1999) is the modular and highly flexible Data Reduction
component whose primary design goals were to simplify data reduction whilst
observing and to provide near-publication quality results. The data reduction
pipeline for SCUBA is based on ORAC-DR, allowing the same approach to be
applied to the processing of archival data, and time-dependent data to be
additionally processed with minimal effort.

ORAC-DR is a recipe driven system that provides prescriptions for all of the
SCUBA observing modes. Since the data are from pointing observations our
adopted recipe is simplified over the generic mapping case in that it can be
assumed that a source is present in the observation, it is close to the centre
of the array and is point-like. Specifically, each pointing observation was
processed in the following way: (i) correction for instrument flat-field; (ii)
spike removal; (iii) extinction correction; (iv) skynoise removal (see
Jenness, Lightfoot \& Holland 1998); (v) rebinning the bolometer data onto a
regular grid; (vi) fitting and removal of any residual image gradient; (vii)
aperture photometry; (viii) flux calibration. These steps are all fully
described in Jenness et al. (2001) but for the purposes of this paper, the
three steps described in the introduction to this section are described in
greater detail below.\\

\noindent (a) Correction for local effects\\

\noindent Pointing observations are always made at the beginning of the night prior to
focusing the telescope and at subsequent times during the night before
refocussing. Therefore, there is a clear potential, indeed certainty, for some
of the pointing maps in the archive to be out-of-focus. There are two
solutions to this problem, one is to ignore all pointing observations taken
immediately before a focus, the second is to ensure that the data extraction
technique is relatively insensitive to out-of-focus conditions. We have taken
the latter course of action because it is also the case that the telescope may
have drifted out of focus during the night due to insufficient focus attention
by the observer. The solution is to use what is equivalent to aperture
photometry. For all observations we use an aperture of 40 arcsec centred
on the source which removes all but the most extreme of the out-of-focus
conditions (see section \ref{sect:post}).\\

\noindent (b) Extinction correction \\

\noindent Observations at 850\mum\ are rather sensitive to the extinction
correction given by S(z)$\propto\exp^{-\tau\sec(z)}$, where $\tau$ is the
 zenith opacity and ranges from abo ut 0.12 to 0.6 for our data set.
 Therefore, as the first step in calibration, great care needs to be taken in
 assigning the correct extinction value to the data.
 
 The extinction for all nights in the SCUBA archive has been determined from a
 combination of SCUBA skydips and the CSO quasi-continuous extinction
 measurements at 225 GHz ($\sim$1.3~mm; see Jenness et al. 2001 and Archibald
 et al. 2001). The resulting data are then fitted by a polynomial of varying
 degree depending on the variation during the night. When the atmosphere is
 very unstable the polynomial fits fail. The astronomical data are
 windowed-out during this period and no extinction values are available for
 the pipeline to attempt a data reduction. While this removes potential data,
 more importantly it limits the inaccuracy of using poorly known extinction
 corrections during periods of large transmission variability.

However, during the early investigations of the accuracy of the pipeline
reduction process, incorrect assignments of the extinction were still found to
be present. On further inspection it was found that this only appeared to
occur close to the time when the atmosphere was in a period of instability,
adjacent to the windowed-out data.  One explanation is that the CSO monitor
operates at a fixed azimuth, and so a pointing observation made just before,
or soon after a period of instability, could be at an azimuth where the
extinction is not well represented by the CSO determination. There are a
number of ways to attempt to compensate for this. The simplest is to increase
the width of the `no extinction correction possible' window. While this would
increase the quality of the remaining data, it would also remove a substantial
dataset from the archive. The second solution is the most complex but
potentially most elegant. This is to attempt to understand the weather pattern
above Mauna Kea from satellite images of the water vapour and model the
results with a time-azimuthally dependent extinction term and to feed this
into the pipeline so that the azimuth of the source at the time can be
modelled against the variable extinction. The third method is the pragmatic
solution, which looks for all such potential problem times and treats the data
by hand on a case-by-case basis.

We have rejected the widening of the data-rejection window, while effort is
not available to pursue the second method. Therefore, we have opted for the
pragmatic solution and have inspected all the data by eye for obvious
problems. Where these are found the data-point is removed rather than an
attempt at a correction made.

When the pipeline is released with the archive the user will be able to make
an assessment of the quality of the night. It is clear that using the data
reduction pipeline without taking care to inspect the extinction pattern for
the night can lead to serious errors in flux determination.\\

\noindent (c) Calibration \\

\noindent Calibration of images (or maps) at submillimetre wavelengths is
best accomplished with observations of point-like sources of known flux, made
close in time to those of the target source. However, our automated procedure
demands that we take a more global approach. Calibration is based on long-term
observations of the two sources CRL618 and Uranus, thereby producing a
time-dependent flux calibration factor (FCF) for SCUBA. These have been very
stable over the period, changes due to upgrades of the instrument are clearly
seen and reflected in the changing FCF. Using this technique an overall
accuracy of around 5 per cent is obtained (Jenness et al.\ 2001).

\section{Post pipeline processing}
\label{sect:post}

The output data from the pipeline for any source are first averaged over an
individual night; there has been no attempt to determine variability within a
single night. The nightly averaged data are first viewed to determine whether
there are any obviously erroneous points. While this is easy to accomplish for
a calibration source, or a source that is not variable, for these variable
extragalactic radio-sources, this introduces a level of subjectivity. For
example, when SCUBA polarimetry is undertaken there is no flag in the
data-header indicating that the polarimeter was in place. This leads to a drop
in the flux by a factor of about two. Although attempts have been made to
retrospectively `flag' all such data in the archive by hand, it is possible
that there are still some `unflagged' data-points remaining. Indeed, during
this work this was discovered for a few data-points, which were easily spotted
and removed.

Two further reasons why individual data-points may be erroneous are that the
extinction has still been incorrectly attributed, or, there were problems with
the SCUBA array. For the latter case, inspection of the individual data images
reveals any obvious problems such as noisy pixels. These data-points are also
removed. Neither of these factors were a common occurrence.

Finally, heating of the dish during early evening can cause notable variations
in the derived flux calibration factor (FCF), which can remain unstable until
late in the evening. Such excursions affect some 5--10 per cent of the data,
and in the most extreme cases can increase the FCF by 50 per cent from its
nominal value, even though the signal is integrated over a 40 arcsec diameter
aperture. During these periods a significant amount of the received flux is
removed from the main beam and spread out into the error lobes. However,
because the resulting images are essentially out of focus, the worst cases are
easily removed by visual inspection of the maps.

\section{Data uncertainties}

In determining the overall uncertainty of the measurement, the flux
calibration uncertainty needs to be added in quadrature to that derived from
the signal-to-noise ratio of the observation. In principle, if multiple
observations of the same source are made on any one night then the
signal-to-noise can be calculated from the scatter in the signals - as is done
for photometry and under the assumption of a non-variable source. This
approach would have the added benefit of including the dish-induced errors for
any given data-set, allowing the use of the standard 5 per cent flux
calibration uncertainty discussed above. In practice, however, although a
source is often observed more than once during a night, not enough data-points
are available to calculate a reliable standard error.

We have thus proceeded as follows. A conservative flux calibration uncertainty
of 10 per cent is adopted. This is based on the scatter of the FCF values
calculated from the CRL618 and Uranus observations, including the `bad'
data-points described above. Because of the need for an automated approach we
chose to split the data into flux bands and then use a typical signal-to-noise
towards the lower end of the range of those calculated for the individual
maps. Our total uncertainties are thus: 10 per cent for fluxes $> 500$~mJy; 12
per cent for 400--500 mJy; 13 per cent for 300--399 mJy; 14 per cent for
200--299 mJy and 17 per cent for 100--199 mJy. Note that the signal-to-noise
for any individual map is $> 5 \sigma$, and for the majority of the data set
is large enough to be negligible in comparison to the 10 per cent flux
calibration uncertainty.

\begin{table}
\caption{The sources observed for this paper along with the date of the
first and last observation and the number of nights during that period on
which the source was observed.}

\begin{tabular}{llccc}
     & & First Obs.   &  Last Obs.    &  Number \\
     & &       &         &   of nights \\
   \hline
0003$-$066 &  & 19970808   &  19991012   &  27 \\
0048$-$097 &  & 19970909   &  19970909   &  1 \\
0106+013 &  & 19970704   &  19991206   &  23 \\
0133+476 &  & 19970404   &  20000420   &  53 \\
0149+218 &  & 19980228   &  19981028   &  2 \\
0215+015 &  & 19971208   &  19971210   &  3 \\
0219+428 &(3C 66A)  & 19970910   &  19970910   &  1 \\
0221+735 &  & 19970809   &  19990217   &  15 \\
0224+671 &  & 19970914   &  19980630   &  3 \\
0234+285 &  & 19971006   &  19991207   &  14 \\
0235+164 &  & 19970713   &  19990217   &  10 \\
0316+413 &(3C 84)  & 19970815   &  20000321   &  53 \\
0336+102 &  & 19970702   &  20000322   &  48 \\
0355+508 &  & 19980105   &  20000304   &  8 \\
0415+379 &(3C 111)  & 19970706   &  20000421   &  38 \\
0420$-$014 &  & 19970810   &  20000409   &  64 \\
0430+052 &(3C 120) & 19970918   &  19981229   &  5 \\
0458$-$020 &  & 19980215   &  19980215   &  1 \\
0521$-$365 &  & 19970920   &  20000422   &  14 \\
0528+134 &  & 19970401   &  20000329   &  47 \\
0529+075 &  & 19971217   &  19981130   &  5 \\
0537$-$441 &  & 19970921   &  19990306   &  8 \\
0552+398 &  & 19970909   &  20000421   &  20 \\
0605$-$085 &  & 19970908   &  20000422   &  13 \\
0607$-$157 &  & 19970409   &  20000418   &  24 \\
0642+449 &  & 19970410   &  19991208   &  25 \\
0716+714 &  & 19970910   &  20000421   &  13 \\
0727$-$115 &  & 19970912   &  20000322   &  12 \\
0735+178 &  & 19971024   &  20000317   &  10 \\
0736+017 &  & 19970410   &  20000421   &  20 \\
0745+241 &  & 19970410   &  20000421   &  20 \\
0754+100 &  & 19970410   &  19980412   &  2 \\
0829+046 &  & 19970410   &  20000205   &  8 \\
0836+710 &  & 19970410   &  20000408   &  18 \\
0851+202 &(OJ 287)  & 19970410   &  20000329   &  25 \\
0917+449 &  & 19970410   &  20000322   &  8 \\
0923+392 &  & 19970404   &  20000422   &  204 \\
0954+685 &  & 19970410   &  20000421   &  31 \\
1034$-$293 &  & 19970410   &  20000421   &  16 \\
1044+719 &  & 19970410   &  20000422   &  9 \\
1055+018 &  & 19970410   &  20000405   &  43 \\
1147+245 &  & 19980130   &  20000405   &  11 \\
1156+295 &  & 19970410   &  20000128   &  29 \\
1213$-$172 &  & 19970410   &  19980319   &  2 \\
1219+285 &  & 19970409   &  20000422   &  39 \\
1226+023 &(3C 273)  & 19970401   &  20000421   &  121 \\
1253$-$055 &(3C 279)  & 19970407   &  20000413   &  116 \\
1308+326 &  & 19970410   &  20000422   &  150 \\
1313$-$333 &  & 19970410   &  19990513   &  9 \\
1334$-$127 &  & 19970408   &  20000331   &  27 \\
1413+135 &  & 19970406   &  20000422   &  33 \\
1418+546 &  & 19970410   &  20000422   &  55 \\
1510$-$089 &  & 19970405   &  20000407   &  27 \\
1514$-$241 &  & 19970405   &  20000317   &  23 \\
1538+149 &  & 19970706   &  20000317   &  5 \\
1548+056 &  & 19980219   &  20000317   &  5 \\
1606+106 &  & 19970405   &  20000317   &  7 \\
1611+343 &  & 19970405   &  20000404   &  93 \\
1633+382 &  & 19970405   &  20000331   &  34 \\
1641+399 &(3C 345)  & 19970410   &  20000329   &  61 \\
\end{tabular}

\label{tab:data}
\end{table}

\begin{table}
\contcaption{ }

\begin{tabular}{llccc}
     & & First Obs.   &  Last Obs.    &  Number\\
     & &    &     &   of nights \\
   \hline
1657$-$262 &  & 19970404   &  20000317   &  11 \\
1730$-$130 &  & 19970404   &  20000403   &  37 \\
1739+552 &  & 19970405   &  20000325   &  19 \\
1741+096 &  & 19970405   &  20000404   & 10\\
1749+096 &  & 19970405   &  20000422   &  21 \\
1803+784 &  & 19970405   &  19990513   &  7 \\
1823+568 &  & 19970405   &  20000322   &  36 \\
1908$-$202 &  & 19970405   &  19990703   &  12 \\
1921$-$293 &  & 19970405   &  20000412   &  24 \\
1923+210 &  & 19981027   &  20000317   &  2 \\
1928+738 &  & 19971005   &  20000325   &  4 \\
1958$-$179 &  & 19970701   &  20000323   &  9 \\
2005+403 &  & 20000317   &  20000317   &  1 \\
2007+776 &  & 19970701   &  19990703   &  12 \\
2021+317 &  & 19970405   &  19971218   &  6 \\
2037+511 &  & 19970405   &  19970405   &  1 \\
2059+034 &  & 19980422   &  19990701   &  2 \\
2145+004 &  & 19970713   &  19990908   &  36 \\
2155$-$304 &  & 19980528   &  19980528   &  1 \\
2155$-$152 &  & 19990703   &  20000409   &  2 \\
2200+420 &(BL Lac) & 19970404   &  19991208   &  65 \\
2201+315 &  & 19970405   &  19971210   &  3 \\
2223$-$052 &(3C 446)  & 19970703   &  20000407   &  36 \\
2227$-$088 &  & 19970920   &  19970920   &  1 \\
2230+114 &  & 19970528   &  19990703   &  7 \\
2251+158 &(3C 454.3) & 19970606   &  19990908   &  35 \\
2255$-$282 &  & 19970808   &  19981230   &  14 \\
2318+049 &  & 19970920   &  19991017   &  19 \\
2345$-$167 &  & 19971005   &  19990703   &  4 \\
\hline
\end{tabular}

\label{tab:data2}
\end{table}

\section{The data}

The observed sources are presented in Table \ref{tab:data} which, for each
source, lists the date of the first and last observation and the number of
nights on which the source was observed. Owing to space constraints the flux
density measurements are not listed and are available in the electronic
version of this paper or from the Centre de Donn\'{e}es astronomiques de
Strasbourg (CDS). Table \ref{tab:eg} provides a subset of the data as an
example. The lightcurves for the best-sampled sources are
presented in Fig. \ref{light-curves1}.

\section{Conclusion}

Due to the nature of the telescope and observing techniques, regular pointing
observations are undertaken by the JCMT as part of the normal observing
pattern. The pointing targets are mainly flat-spectrum radio sources. The
resulting data are a potentially valuable resource for monitoring studies,
especially multifrequency studies of active galaxies looking for links between
the various spectral signatures of primary and secondary emission
mechanisms. However, before they can be useful the data must be reliably
calibrated. We have shown that with care, the data can be calibrated to an
accuracy that is extremely useful. Furthermore, the development and refinement
of the pipeline process means that it is a relatively simple task to extract
and calibrate all these data and make them available to the scientific
community. This is the first paper in what will be an ongoing release of data
that can be used in a variety of studies.

\begin{table}
\caption{Results for two of the sources. The full results are available in the
  electronic form of this paper. Modified Julian Date (MJD) is defined as
  Julian Date $-$ 2400000.5.}
\label{tab:eg}
\begin{center}
\begin{tabular}{ccrr}
Date  & Date & Flux  & Error \\
/ MJD & / UT & / mJy & / mJy  \\ \hline       
\multicolumn{4}{c}{$1741+096$} \\
51638 &20000404& 1181.1&  35.2\\
51620 &20000317& 1139.1& 113.9\\
51311 &19990513&  740.5&  74.0\\
50975 &19980611& 1585.3& 158.5\\
50635 &19970706& 1741.5& 174.2\\
50630 &19970701& 1899.1& 189.9\\
50543 &19970405& 1719.3& 171.9\\
\multicolumn{4}{c}{$2059+034$} \\
51360& 19990701&   387.7 &  50.4 \\
50925& 19980422&   577.8 &  57.8 \\

\hline
\end{tabular}
\end{center}

\end{table}

\begin{figure*}

\begin{tabular}{cc}
\includegraphics[width=78mm]{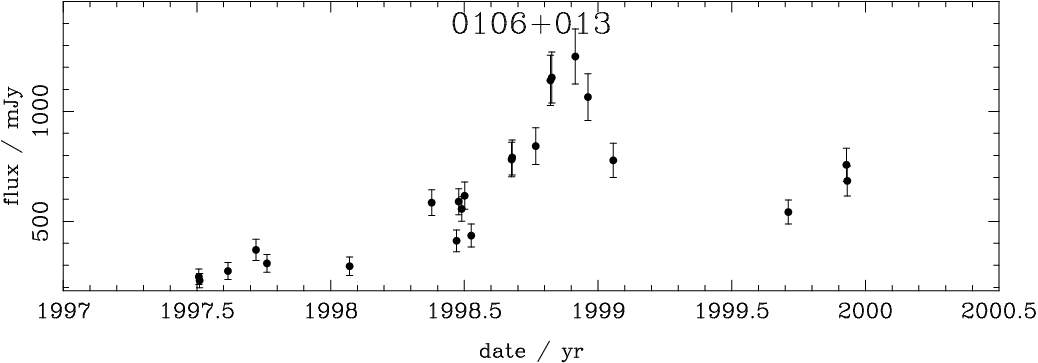}
&
\includegraphics[width=78mm]{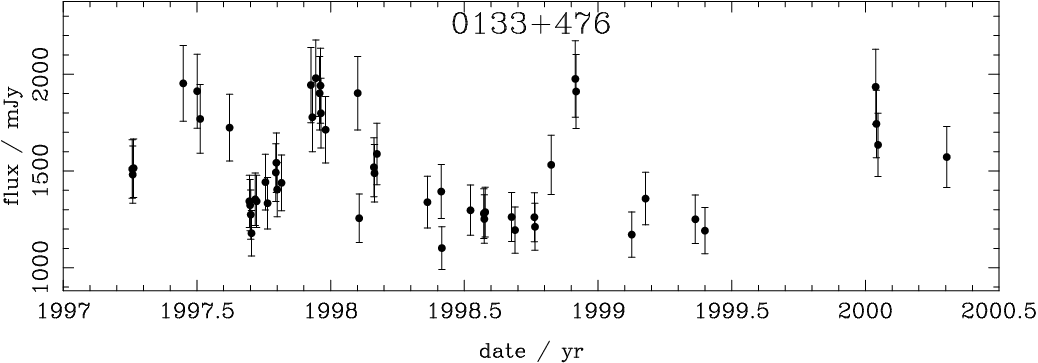}
\\
\includegraphics[width=78mm]{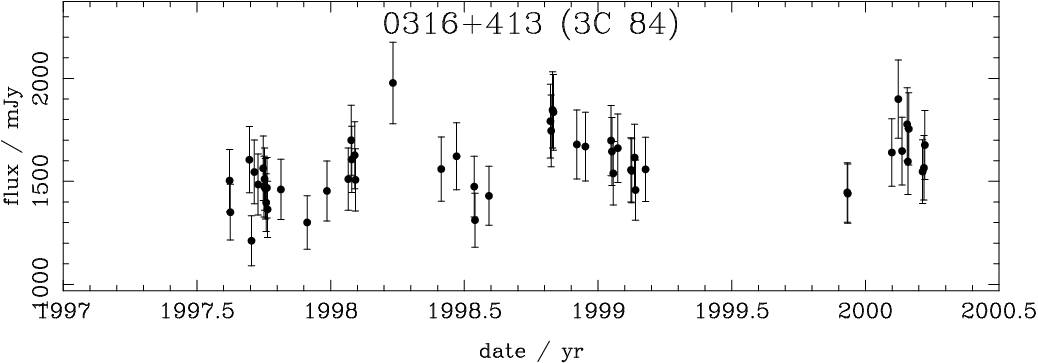}
&
\includegraphics[width=78mm]{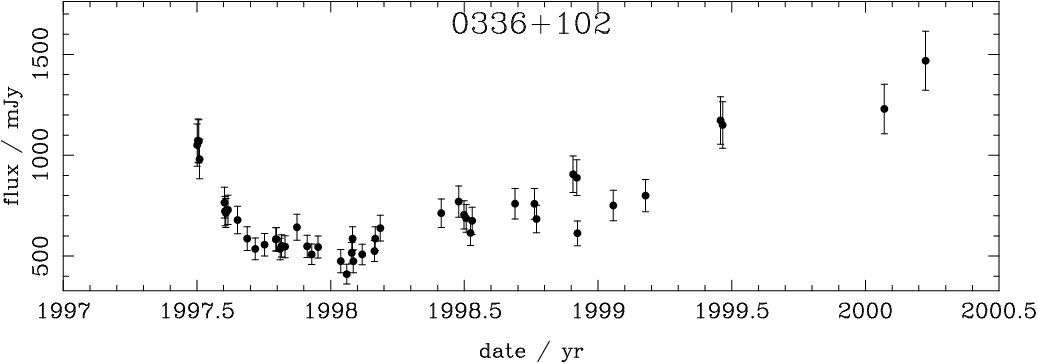}
\\
\includegraphics[width=78mm]{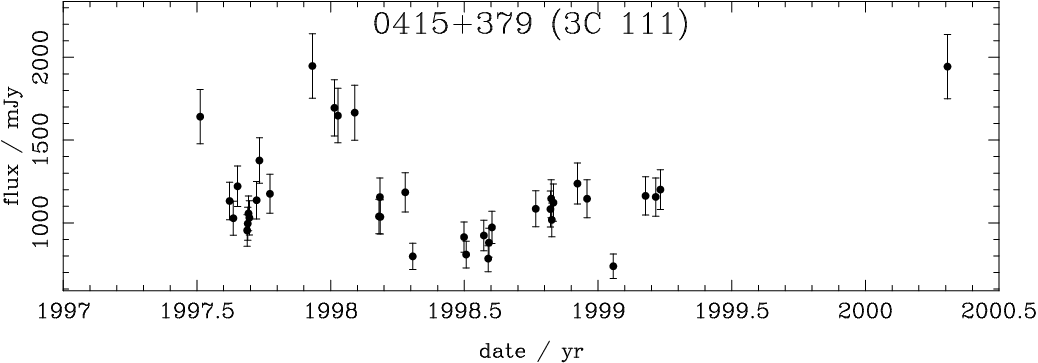}
&
\includegraphics[width=78mm]{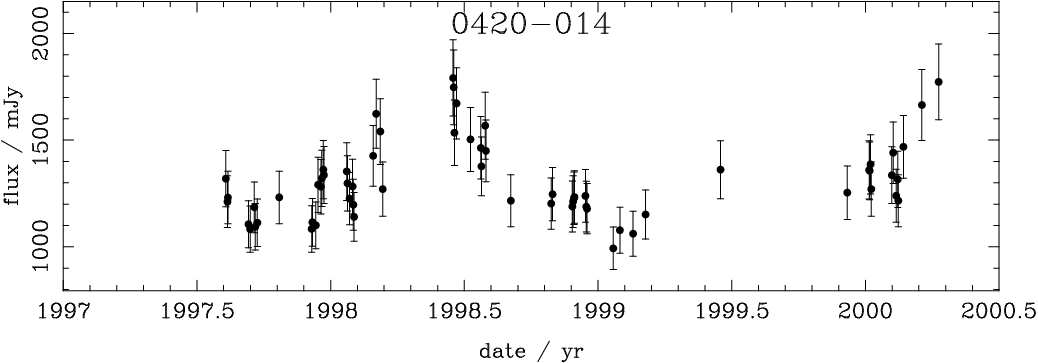}
\\
\includegraphics[width=78mm]{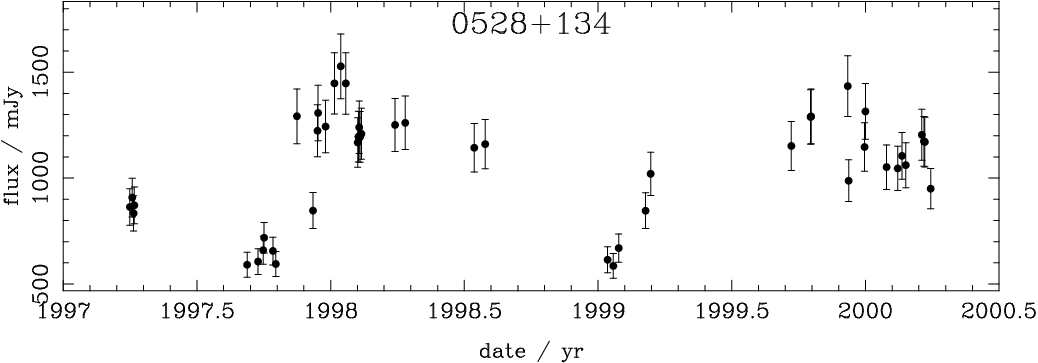}
&
\includegraphics[width=78mm]{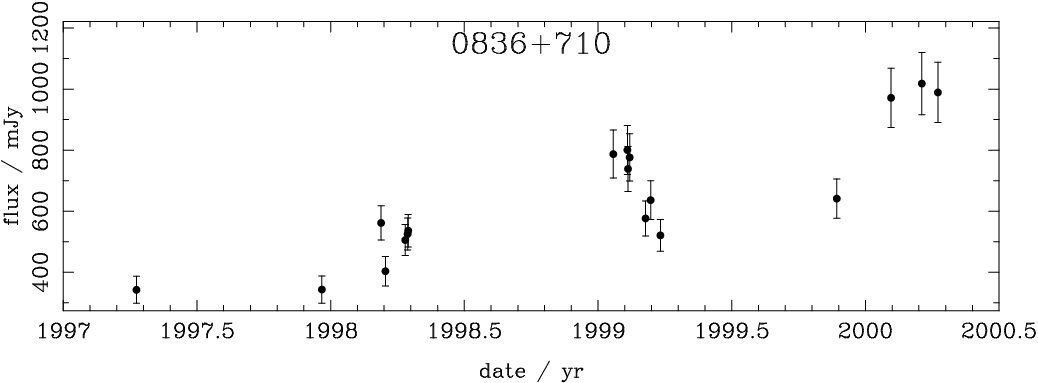}
\\
\includegraphics[width=78mm]{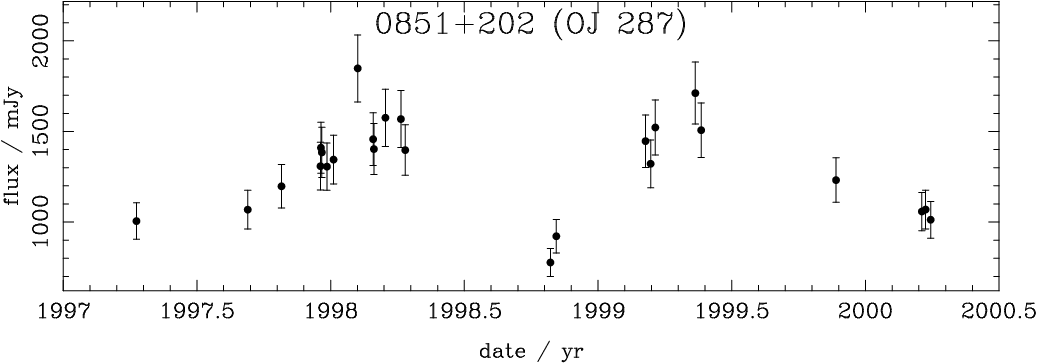}
&
\includegraphics[width=78mm]{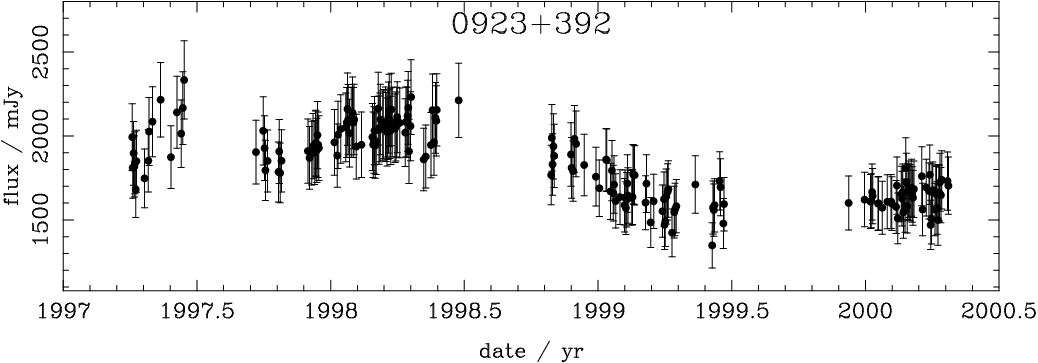}
\\
\includegraphics[width=78mm]{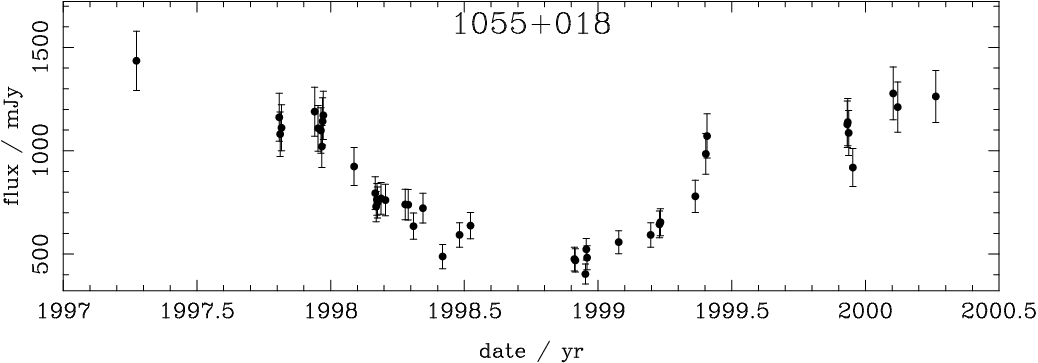}
&
\includegraphics[width=78mm]{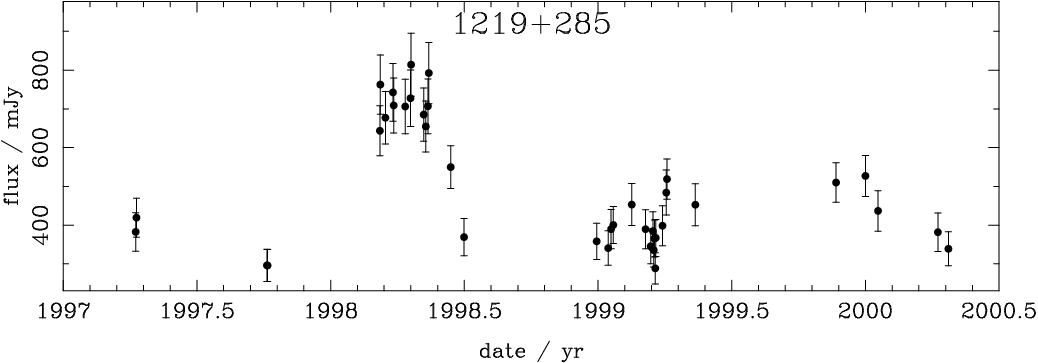}
\\
\includegraphics[width=78mm]{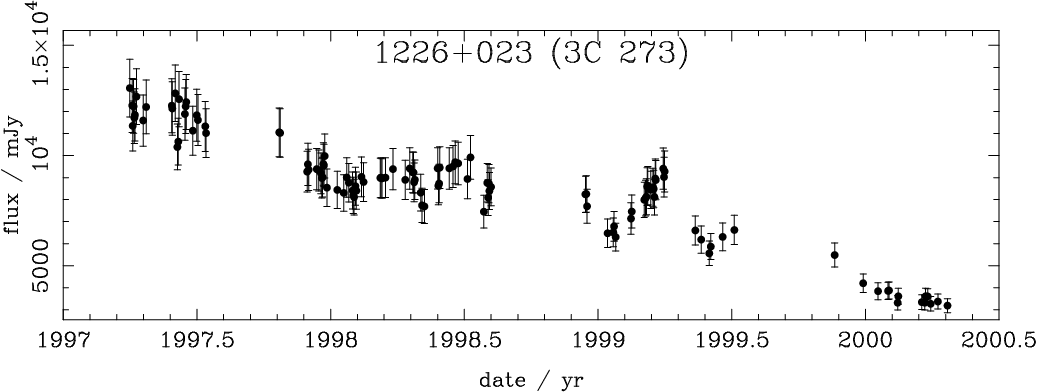}
&
\includegraphics[width=78mm]{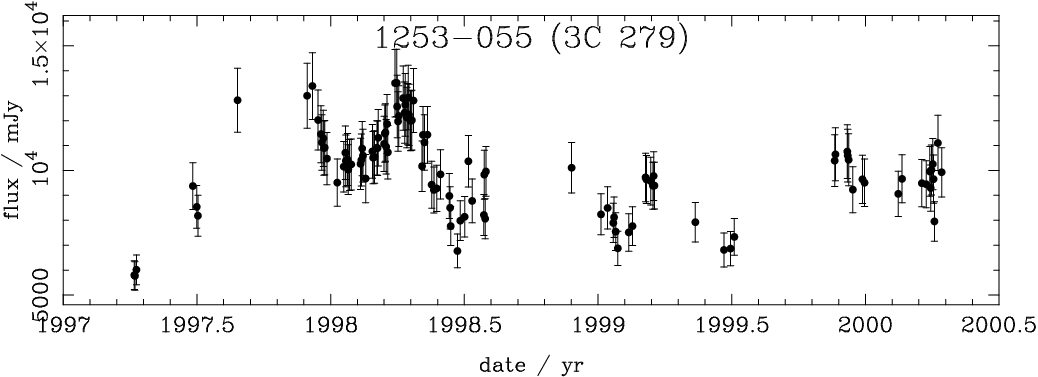}
\\
\includegraphics[width=78mm]{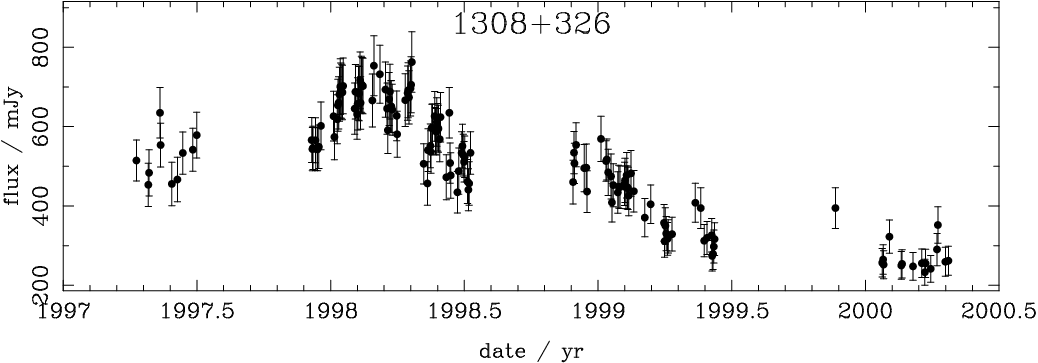}
&
\includegraphics[width=78mm]{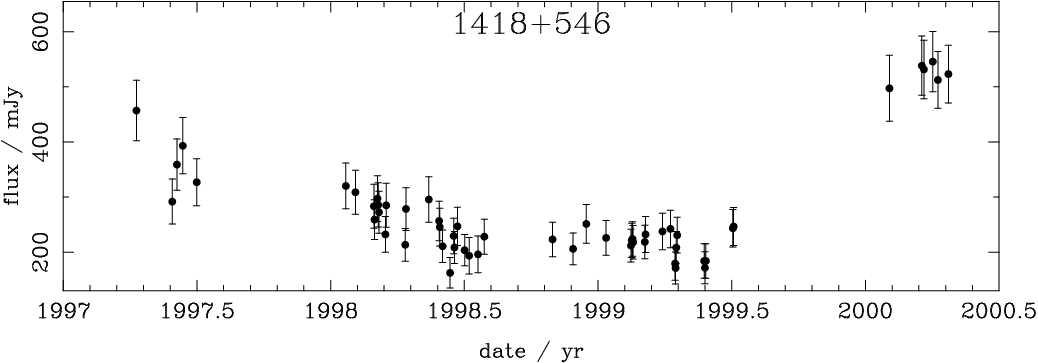}
\\

\end{tabular}

\caption{Light curves for well-sampled data sets.}
\label{light-curves1}

\end{figure*}

\begin{figure*}

\begin{tabular}{cc}

\includegraphics[width=78mm]{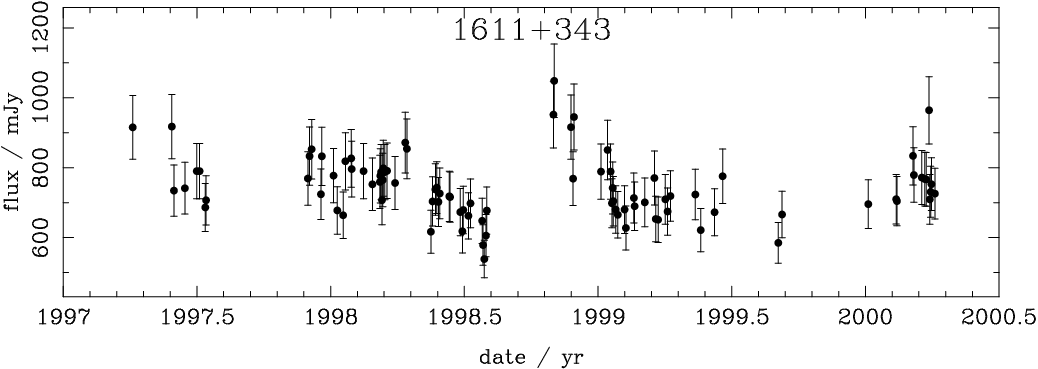}
&
\includegraphics[width=78mm]{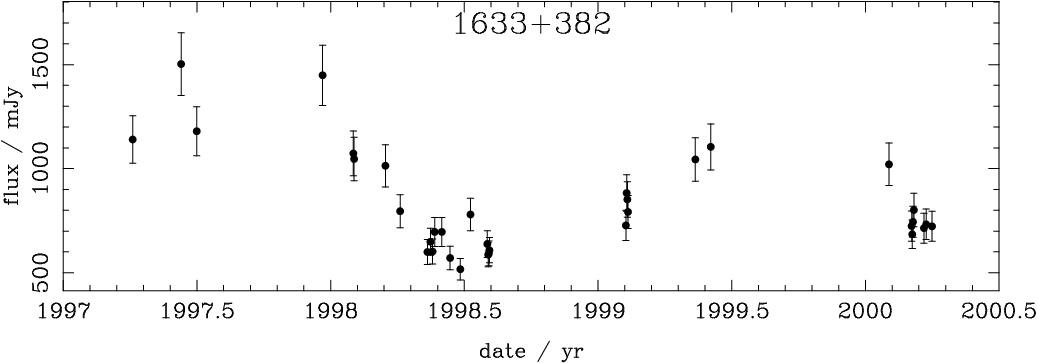}
\\
\includegraphics[width=78mm]{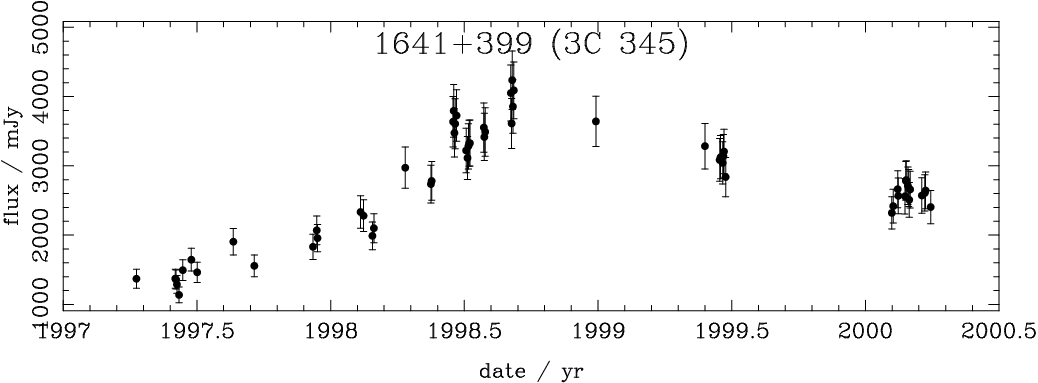}
&
\includegraphics[width=78mm]{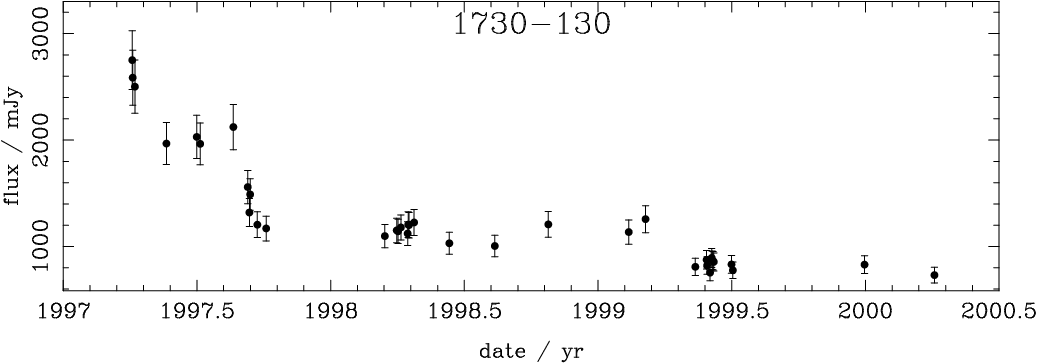}
\\
\includegraphics[width=78mm]{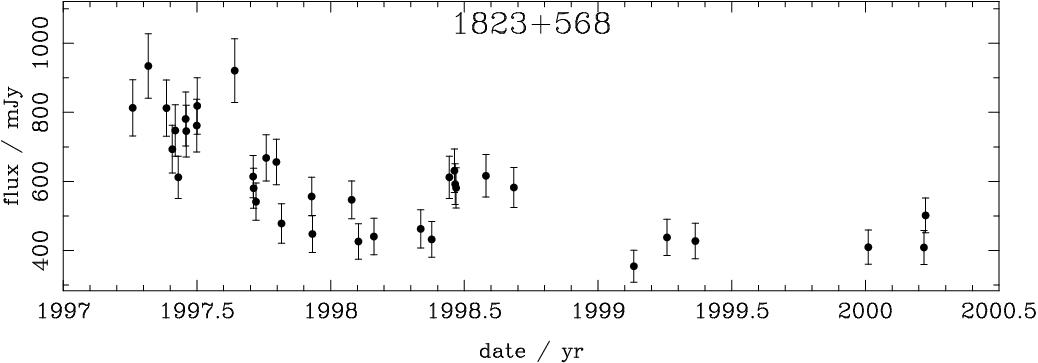}
&
\includegraphics[width=78mm]{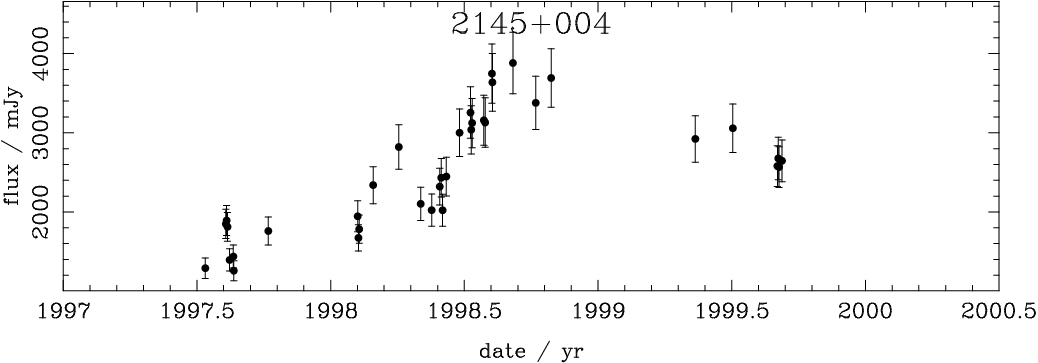}
\\
\includegraphics[width=78mm]{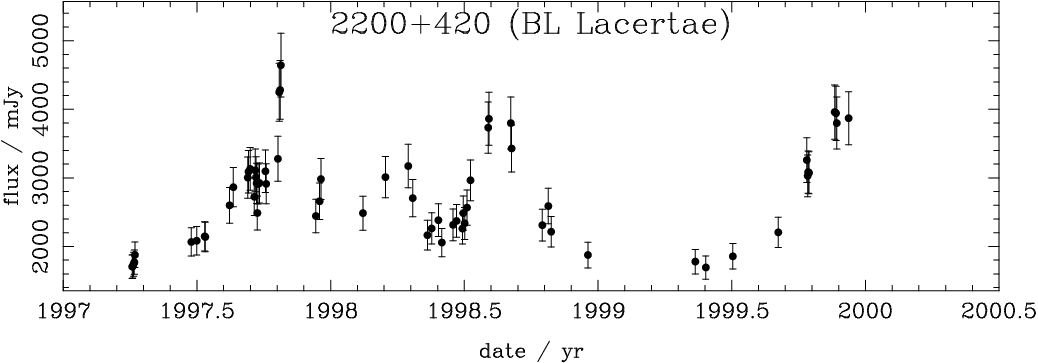}
&
\includegraphics[width=78mm]{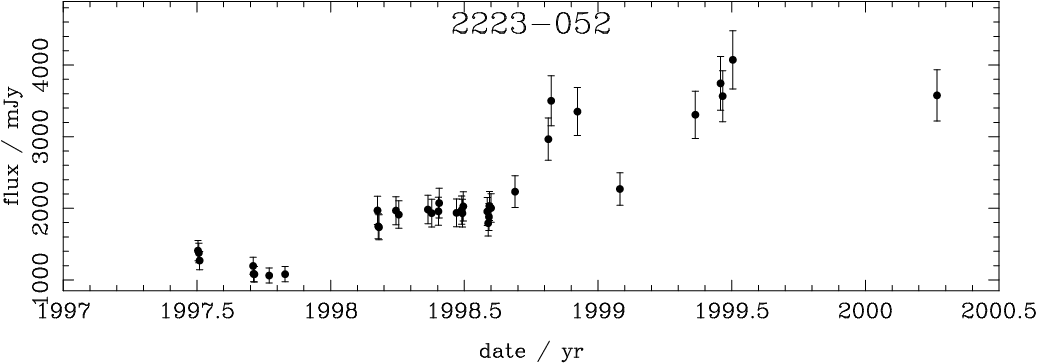}
\\

\multicolumn{2}{c}{ %
\includegraphics[width=78mm]{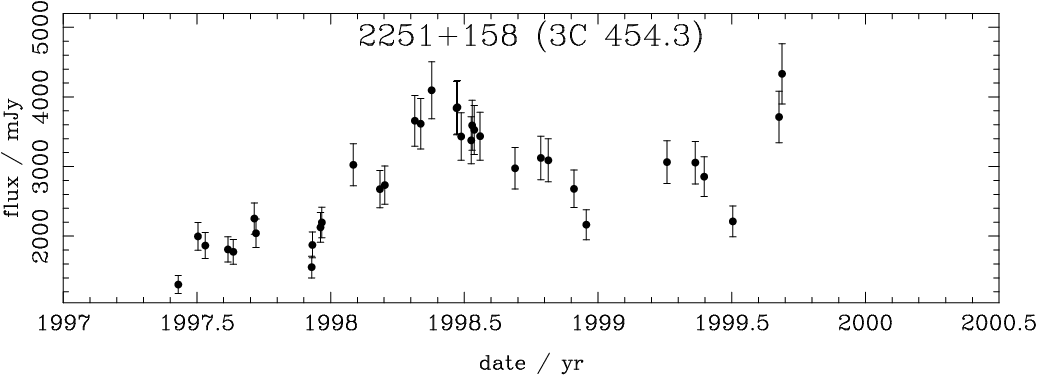}
}

\end{tabular}

\contcaption{ }
\end{figure*}


\section*{ACKNOWLEDGMENTS}

The James Clerk Maxwell Telescope is operated by the Joint Astronomy Centre in
Hilo, Hawaii on behalf of the parent organisations PPARC in the United Kingdom,
the National Research Council of Canada and The Netherlands Organisation for
Scientific Research. J.A.S.\ acknowledges support from PPARC.

We acknowledge the support software provided by the Starlink Project which is
run by CCLRC on behalf of PPARC.

\bsp

\clearpage

\clearpage
\begin{table*}
\contcaption{ }
\begin{tabular}{rrrrrrrrrrrr}
Date & Date & Flux& Error&Date & Date & Flux& Error&Date & Date & Flux& Error\\
MJD  &  UT  & / mJy & / mJy&MJD  &  UT  & / mJy & / mJy&MJD  &  UT  & / mJy & / mJy\\\hline
\hline\multicolumn{12}{c}{\textit{0003-066}}\\
\hline \textbf{51463} & 19991012 & 1198.9 & 119 & \textbf{51010} & 19980716 & 920.0 & 92 & \textbf{50961} & 19980528 & 1059.8 & 106\\
\textbf{51118} & 19981101 & 1407.6 & 140 & \textbf{51006} & 19980712 & 1046.6 & 104 & \textbf{50960} & 19980527 & 1002.2 & 100\\
\textbf{51116} & 19981030 & 1501.2 & 150 & \textbf{51004} & 19980710 & 1056.2 & 105 & \textbf{50951} & 19980518 & 982.7 & 98\\
\textbf{51114} & 19981028 & 1383.5 & 138 & \textbf{50999} & 19980705 & 944.2 & 94 & \textbf{50945} & 19980512 & 909.9 & 91\\
\textbf{51059} & 19980903 & 1069.7 & 107 & \textbf{50995} & 19980701 & 1044.8 & 104 & \textbf{50922} & 19980419 & 1120.1 & 112\\
\end{tabular}
\end{table*}
\clearpage
\begin{table*}
\contcaption{ }
\begin{tabular}{rrrrrrrrrrrr}
Date & Date & Flux& Error&Date & Date & Flux& Error&Date & Date & Flux& Error\\
MJD  &  UT  & / mJy & / mJy&MJD  &  UT  & / mJy & / mJy&MJD  &  UT  & / mJy & / mJy\\\hline
\textbf{51199} & 19990121 & 409.5 & 49 & \textbf{50880} & 19980308 & 238.4 & 33 & \textbf{50751} & 19971030 & 214.7 & 30\\
\textbf{51147} & 19981130 & 479.9 & 57 & \textbf{50839} & 19980126 & 197.0 & 33 & \textbf{50723} & 19971002 & 206.1 & 28\\
\textbf{51114} & 19981028 & 391.9 & 51 & \textbf{50804} & 19971222 & 237.5 & 33 & \textbf{50711} & 19970920 & 200.0 & 28\\
\textbf{51022} & 19980728 & 202.9 & 28 & \textbf{50790} & 19971208 & 182.4 & 31 & \textbf{50669} & 19970809 & 152.5 & 25\\
\hline\multicolumn{12}{c}{\textit{0224+671}}\\
\hline \textbf{50994} & 19980630 & 666.8 & 66 & \textbf{50859} & 19980215 & 670.1 & 67 & \textbf{50705} & 19970914 & 665.0 & 66\\
\end{tabular}
\end{table*}
\clearpage
\begin{table*}
\contcaption{ }
\begin{tabular}{rrrrrrrrrrrr}
Date & Date & Flux& Error&Date & Date & Flux& Error&Date & Date & Flux& Error\\
MJD  &  UT  & / mJy & / mJy&MJD  &  UT  & / mJy & / mJy&MJD  &  UT  & / mJy & / mJy\\\hline
\textbf{51004} & 19980710 & 614.5 & 61 & \textbf{50781} & 19971129 & 548.2 & 54 & \textbf{50633} & 19970704 & 1069.0 & 106\\
\textbf{50998} & 19980704 & 687.4 & 68 & \textbf{50767} & 19971115 & 642.8 & 64 & \textbf{50632} & 19970703 & 1072.6 & 107\\
\textbf{50995} & 19980701 & 704.4 & 70 & \textbf{50751} & 19971030 & 547.0 & 54 & \textbf{50631} & 19970702 & 1050.6 & 105\\
\hline\multicolumn{12}{c}{\textit{0355+508}}\\
\hline \textbf{51607} & 20000304 & 1445.7 & 144 & \textbf{51243} & 19990306 & 1707.9 & 170 & \textbf{50828} & 19980115 & 1516.6 & 151\\
\textbf{51606} & 20000303 & 1754.9 & 175 & \textbf{51199} & 19990121 & 1415.8 & 141 & \textbf{50818} & 19980105 & 1628.1 & 162\\
\end{tabular}
\end{table*}
\clearpage
\begin{table*}
\contcaption{ }
\begin{tabular}{rrrrrrrrrrrr}
Date & Date & Flux& Error&Date & Date & Flux& Error&Date & Date & Flux& Error\\
MJD  &  UT  & / mJy & / mJy&MJD  &  UT  & / mJy & / mJy&MJD  &  UT  & / mJy & / mJy\\\hline
\textbf{51640} & 20000406 & 2188.8 & 218 & \textbf{50898} & 19980326 & 3133.4 & 313 & \textbf{50745} & 19971024 & 1977.4 & 197\\
\textbf{51620} & 20000317 & 2454.3 & 245 & \textbf{50878} & 19980306 & 3189.5 & 319 & \textbf{50734} & 19971013 & 2004.0 & 200\\
\textbf{51243} & 19990306 & 2708.6 & 270 & \textbf{50877} & 19980305 & 2657.9 & 265 & \textbf{50711} & 19970920 & 2158.7 & 215\\
\textbf{51199} & 19990121 & 2344.7 & 234 & \textbf{50870} & 19980226 & 2424.6 & 242 &  &  &  & \\
\hline\multicolumn{12}{c}{\textit{0528+134}}\\
\hline \textbf{51632} & 20000329 & 950.0 & 95 & \textbf{51243} & 19990306 & 846.1 & 84 & \textbf{50806} & 19971224 & 1243.4 & 124\\
\end{tabular}
\end{table*}
\clearpage
\begin{table*}
\contcaption{ }
\begin{tabular}{rrrrrrrrrrrr}
Date & Date & Flux& Error&Date & Date & Flux& Error&Date & Date & Flux& Error\\
MJD  &  UT  & / mJy & / mJy&MJD  &  UT  & / mJy & / mJy&MJD  &  UT  & / mJy & / mJy\\\hline
\textbf{51223} & 19990214 & 1418.1 & 141 & \textbf{50915} & 19980412 & 1983.1 & 198 & \textbf{50548} & 19970410 & 1638.9 & 163\\
\textbf{51205} & 19990127 & 1161.0 & 116 & \textbf{50888} & 19980316 & 2123.1 & 212 & \textbf{50547} & 19970409 & 1565.7 & 156\\
\hline\multicolumn{12}{c}{\textit{0642+449}}\\
\hline \textbf{51520} & 19991208 & 531.9 & 53 & \textbf{51020} & 19980726 & 565.1 & 56 & \textbf{50789} & 19971207 & 602.9 & 60\\
\textbf{51243} & 19990306 & 441.9 & 53 & \textbf{50914} & 19980411 & 435.8 & 52 & \textbf{50746} & 19971025 & 525.9 & 52\\
\textbf{51199} & 19990121 & 430.3 & 51 & \textbf{50839} & 19980126 & 490.6 & 58 & \textbf{50730} & 19971009 & 474.8 & 57\\
\end{tabular}
\end{table*}
\clearpage
\begin{table*}
\contcaption{ }
\begin{tabular}{rrrrrrrrrrrr}
Date & Date & Flux& Error&Date & Date & Flux& Error&Date & Date & Flux& Error\\
MJD  &  UT  & / mJy & / mJy&MJD  &  UT  & / mJy & / mJy&MJD  &  UT  & / mJy & / mJy\\\hline
\textbf{50915} & 19980412 & 641.6 & 64 & \textbf{50876} & 19980304 & 916.2 & 91 &  &  &  & \\
\hline\multicolumn{12}{c}{\textit{0836+710}}\\
\hline \textbf{51642} & 20000408 & 989.3 & 98 & \textbf{51243} & 19990306 & 576.2 & 57 & \textbf{50918} & 19980415 & 525.3 & 52\\
\textbf{51620} & 20000317 & 1018.0 & 101 & \textbf{51221} & 19990212 & 776.5 & 77 & \textbf{50915} & 19980412 & 505.5 & 50\\
\textbf{51578} & 20000204 & 971.5 & 97 & \textbf{51219} & 19990210 & 738.5 & 73 & \textbf{50888} & 19980316 & 403.3 & 48\\
\textbf{51504} & 19991122 & 641.2 & 64 & \textbf{51218} & 19990209 & 800.4 & 80 & \textbf{50882} & 19980310 & 561.3 & 56\\
\end{tabular}
\end{table*}
\clearpage
\begin{table*}
\contcaption{ }
\begin{tabular}{rrrrrrrrrrrr}
Date & Date & Flux& Error&Date & Date & Flux& Error&Date & Date & Flux& Error\\
MJD  &  UT  & / mJy & / mJy&MJD  &  UT  & / mJy & / mJy&MJD  &  UT  & / mJy & / mJy\\\hline
\textbf{51601} & 20000227 & 1669.7 & 167 & \textbf{51175} & 19981228 & 1757.5 & 175 & \textbf{50823} & 19980110 & 2006.0 & 200\\
\textbf{51600} & 20000226 & 1580.7 & 158 & \textbf{51159} & 19981212 & 1826.9 & 182 & \textbf{50822} & 19980109 & 1883.3 & 188\\
\textbf{51599} & 20000225 & 1723.9 & 172 & \textbf{51148} & 19981201 & 1952.9 & 195 & \textbf{50818} & 19980105 & 1961.4 & 196\\
\textbf{51598} & 20000224 & 1808.1 & 180 & \textbf{51146} & 19981129 & 1981.8 & 198 & \textbf{50797} & 19971215 & 1927.7 & 192\\
\textbf{51597} & 20000223 & 1584.9 & 158 & \textbf{51144} & 19981127 & 1790.3 & 179 & \textbf{50796} & 19971214 & 1947.2 & 194\\
\textbf{51595} & 20000221 & 1545.2 & 154 & \textbf{51142} & 19981125 & 1810.2 & 181 & \textbf{50795} & 19971213 & 2004.8 & 200\\
\end{tabular}
\end{table*}
\clearpage
\begin{table*}
\contcaption{ }
\begin{tabular}{rrrrrrrrrrrr}
Date & Date & Flux& Error&Date & Date & Flux& Error&Date & Date & Flux& Error\\
MJD  &  UT  & / mJy & / mJy&MJD  &  UT  & / mJy & / mJy&MJD  &  UT  & / mJy & / mJy\\\hline
\textbf{51652} & 20000418 & 673.6 & 67 & \textbf{51243} & 19990306 & 799.3 & 79 & \textbf{50791} & 19971209 & 363.7 & 47\\
\textbf{51325} & 19990527 & 515.2 & 51 & \textbf{51205} & 19990127 & 780.4 & 78 & \textbf{50745} & 19971024 & 447.8 & 53\\
\textbf{51311} & 19990513 & 493.1 & 59 & \textbf{50915} & 19980412 & 998.9 & 99 & \textbf{50548} & 19970410 & 766.2 & 76\\
\textbf{51274} & 19990406 & 681.0 & 68 & \textbf{50888} & 19980316 & 842.9 & 84 & \textbf{51250} & 19990313 & 911.1 & 91\\
\hline\multicolumn{12}{c}{\textit{1044+719}}\\
\hline \textbf{51656} & 20000422 & 861.2 & 86 & \textbf{50888} & 19980316 & 549.8 & 55 & \textbf{50799} & 19971217 & 527.8 & 52\\
\end{tabular}
\end{table*}
\clearpage
\begin{table*}
\contcaption{ }
\begin{tabular}{rrrrrrrrrrrr}
Date & Date & Flux& Error&Date & Date & Flux& Error&Date & Date & Flux& Error\\
MJD  &  UT  & / mJy & / mJy&MJD  &  UT  & / mJy & / mJy&MJD  &  UT  & / mJy & / mJy\\\hline
\textbf{51272} & 19990404 & 518.9 & 51 & \textbf{51192} & 19990114 & 340.8 & 44 & \textbf{50888} & 19980316 & 677.2 & 67\\
\textbf{51271} & 19990403 & 484.0 & 58 & \textbf{51176} & 19981229 & 358.1 & 46 & \textbf{50881} & 19980309 & 762.8 & 76\\
\textbf{51266} & 19990329 & 398.5 & 51 & \textbf{50995} & 19980701 & 369.1 & 48 & \textbf{50880} & 19980308 & 643.6 & 64\\
\textbf{51257} & 19990320 & 366.9 & 47 & \textbf{50977} & 19980613 & 549.9 & 55 & \textbf{50727} & 19971006 & 295.7 & 41\\
\textbf{51256} & 19990319 & 288.5 & 40 & \textbf{50947} & 19980514 & 792.4 & 79 & \textbf{50726} & 19971005 & 296.3 & 41\\
\textbf{51255} & 19990318 & 365.5 & 47 & \textbf{50946} & 19980513 & 706.9 & 70 & \textbf{50548} & 19970410 & 419.2 & 50\\
\end{tabular}
\end{table*}
\clearpage
\begin{table*}
\contcaption{ }
\begin{tabular}{rrrrrrrrrrrr}
Date & Date & Flux& Error&Date & Date & Flux& Error&Date & Date & Flux& Error\\
MJD  &  UT  & / mJy & / mJy&MJD  &  UT  & / mJy & / mJy&MJD  &  UT  & / mJy & / mJy\\\hline
\textbf{51636} & 20000402 & 9653.3 & 965 & \textbf{51022} & 19980728 & 8215.0 & 821 & \textbf{50878} & 19980306 & 11311.3 & 1131\\
\textbf{51635} & 20000401 & 10255.0 & 1025 & \textbf{51006} & 19980712 & 8780.3 & 878 & \textbf{50877} & 19980305 & 10893.3 & 1089\\
\textbf{51633} & 20000330 & 10017.7 & 1001 & \textbf{51001} & 19980707 & 10366.5 & 1036 & \textbf{50876} & 19980304 & 10909.8 & 1091\\
\textbf{51632} & 20000329 & 9293.7 & 929 & \textbf{50996} & 19980702 & 8139.3 & 813 & \textbf{50873} & 19980301 & 10542.5 & 1054\\
\textbf{51631} & 20000328 & 9960.1 & 996 & \textbf{50990} & 19980626 & 7986.5 & 798 & \textbf{50871} & 19980227 & 10514.4 & 1051\\
\textbf{51626} & 20000323 & 9448.9 & 944 & \textbf{50986} & 19980622 & 6769.7 & 677 & \textbf{50870} & 19980226 & 10764.0 & 1076\\
\end{tabular}
\end{table*}
\clearpage
\begin{table*}
\contcaption{ }
\begin{tabular}{rrrrrrrrrrrr}
Date & Date & Flux& Error&Date & Date & Flux& Error&Date & Date & Flux& Error\\
MJD  &  UT  & / mJy & / mJy&MJD  &  UT  & / mJy & / mJy&MJD  &  UT  & / mJy & / mJy\\\hline
\textbf{51337} & 19990608 & 316.6 & 41 & \textbf{50994} & 19980630 & 528.2 & 52 & \textbf{50851} & 19980207 & 684.4 & 68\\
\textbf{51336} & 19990607 & 297.5 & 41 & \textbf{50993} & 19980629 & 551.0 & 55 & \textbf{50850} & 19980206 & 657.7 & 65\\
\textbf{51335} & 19990606 & 278.9 & 39 & \textbf{50992} & 19980628 & 534.0 & 53 & \textbf{50849} & 19980205 & 631.1 & 63\\
\textbf{51334} & 19990605 & 274.2 & 38 & \textbf{50988} & 19980624 & 487.9 & 58 & \textbf{50847} & 19980203 & 687.9 & 68\\
\textbf{51333} & 19990604 & 326.2 & 42 & \textbf{50986} & 19980622 & 434.6 & 52 & \textbf{50846} & 19980202 & 645.6 & 64\\
\textbf{51327} & 19990529 & 319.4 & 41 & \textbf{50977} & 19980613 & 477.2 & 57 & \textbf{50830} & 19980117 & 702.6 & 70\\
\end{tabular}
\end{table*}
\clearpage
\begin{table*}
\contcaption{ }
\begin{tabular}{rrrrrrrrrrrr}
Date & Date & Flux& Error&Date & Date & Flux& Error&Date & Date & Flux& Error\\
MJD  &  UT  & / mJy & / mJy&MJD  &  UT  & / mJy & / mJy&MJD  &  UT  & / mJy & / mJy\\\hline
\textbf{51621} & 20000318 & 643.8 & 64 & \textbf{50906} & 19980403 & 1370.1 & 137 & \textbf{50787} & 19971205 & 1389.4 & 138\\
\textbf{51620} & 20000317 & 713.3 & 71 & \textbf{50887} & 19980315 & 1435.3 & 143 & \textbf{50632} & 19970703 & 738.0 & 73\\
\textbf{51311} & 19990513 & 706.0 & 70 & \textbf{50873} & 19980301 & 1229.2 & 122 & \textbf{50630} & 19970701 & 742.6 & 74\\
\textbf{51225} & 19990216 & 677.8 & 67 & \textbf{50861} & 19980217 & 1223.4 & 122 & \textbf{50600} & 19970601 & 931.8 & 93\\
\textbf{51023} & 19980729 & 566.9 & 56 & \textbf{50847} & 19980203 & 1270.1 & 127 & \textbf{50597} & 19970529 & 850.5 & 85\\
\textbf{51022} & 19980728 & 976.4 & 97 & \textbf{50845} & 19980201 & 1290.4 & 129 & \textbf{50594} & 19970526 & 508.5 & 50\\
\end{tabular}
\end{table*}
\clearpage
\begin{table*}
\contcaption{ }
\begin{tabular}{rrrrrrrrrrrr}
Date & Date & Flux& Error&Date & Date & Flux& Error&Date & Date & Flux& Error\\
MJD  &  UT  & / mJy & / mJy&MJD  &  UT  & / mJy & / mJy&MJD  &  UT  & / mJy & / mJy\\\hline
\hline\multicolumn{12}{c}{\textit{1548+056}}\\
\hline \textbf{51620} & 20000317 & 705.7 & 70 & \textbf{51360} & 19990701 & 555.2 & 55 & \textbf{50863} & 19980219 & 845.1 & 84\\
\textbf{51588} & 20000214 & 785.2 & 78 & \textbf{50888} & 19980316 & 824.2 & 82 &  &  &  & \\
\hline\multicolumn{12}{c}{\textit{1606+106}}\\
\hline \textbf{51620} & 20000317 & 683.3 & 68 & \textbf{50871} & 19980227 & 375.3 & 48 & \textbf{50543} & 19970405 & 427.7 & 51\\
\textbf{51332} & 19990603 & 565.5 & 56 & \textbf{50870} & 19980226 & 304.7 & 39 & \textbf{51253} & 19990316 & 550.1 & 55\\
\end{tabular}
\end{table*}
\clearpage
\begin{table*}
\contcaption{ }
\begin{tabular}{rrrrrrrrrrrr}
Date & Date & Flux& Error&Date & Date & Flux& Error&Date & Date & Flux& Error\\
MJD  &  UT  & / mJy & / mJy&MJD  &  UT  & / mJy & / mJy&MJD  &  UT  & / mJy & / mJy\\\hline
\hline\multicolumn{12}{c}{\textit{1641+399 (3C 345)}}\\
\hline \textbf{51632} & 20000329 & 2404.6 & 240 & \textbf{51324} & 19990526 & 3283.4 & 328 & \textbf{50950} & 19980517 & 2736.5 & 273\\
\textbf{51625} & 20000322 & 2644.3 & 264 & \textbf{51175} & 19981228 & 3641.7 & 364 & \textbf{50915} & 19980412 & 2973.0 & 297\\
\textbf{51624} & 20000321 & 2609.2 & 260 & \textbf{51063} & 19980907 & 4090.1 & 409 & \textbf{50872} & 19980228 & 2100.1 & 210\\
\textbf{51620} & 20000317 & 2571.4 & 257 & \textbf{51062} & 19980906 & 3856.1 & 385 & \textbf{50870} & 19980226 & 1990.3 & 199\\
\textbf{51604} & 20000301 & 2657.8 & 265 & \textbf{51061} & 19980905 & 4235.5 & 423 & \textbf{50858} & 19980214 & 2280.5 & 228\\
\end{tabular}
\end{table*}
\clearpage
\begin{table*}
\contcaption{ }
\begin{tabular}{rrrrrrrrrrrr}
Date & Date & Flux& Error&Date & Date & Flux& Error&Date & Date & Flux& Error\\
MJD  &  UT  & / mJy & / mJy&MJD  &  UT  & / mJy & / mJy&MJD  &  UT  & / mJy & / mJy\\\hline
\hline\multicolumn{12}{c}{\textit{1749+096}}\\
\hline \textbf{51656} & 20000422 & 1256.1 & 125 & \textbf{50962} & 19980529 & 2273.4 & 227 & \textbf{50710} & 19970919 & 1940.8 & 194\\
\textbf{51636} & 20000402 & 1098.1 & 109 & \textbf{50961} & 19980528 & 2283.5 & 228 & \textbf{50630} & 19970701 & 1585.5 & 158\\
\textbf{51620} & 20000317 & 1122.1 & 112 & \textbf{50960} & 19980527 & 2280.5 & 228 & \textbf{50605} & 19970606 & 1101.8 & 110\\
\textbf{51311} & 19990513 & 1276.1 & 127 & \textbf{50902} & 19980330 & 2197.1 & 219 & \textbf{50604} & 19970605 & 1270.0 & 127\\
\textbf{51271} & 19990403 & 1111.2 & 111 & \textbf{50851} & 19980207 & 1725.2 & 172 & \textbf{50599} & 19970531 & 1284.8 & 128\\
\end{tabular}
\end{table*}
\clearpage
\begin{table*}
\contcaption{ }
\begin{tabular}{rrrrrrrrrrrr}
Date & Date & Flux& Error&Date & Date & Flux& Error&Date & Date & Flux& Error\\
MJD  &  UT  & / mJy & / mJy&MJD  &  UT  & / mJy & / mJy&MJD  &  UT  & / mJy & / mJy\\\hline
\hline\multicolumn{12}{c}{\textit{2005+403}}\\
\hline \textbf{51620} & 20000317 & 249.1 & 34 &  &  &  &  &  &  &  & \\
\hline\multicolumn{12}{c}{\textit{2007+776}}\\
\hline \textbf{51362} & 19990703 & 314.2 & 40 & \textbf{50981} & 19980617 & 427.5 & 51 & \textbf{50918} & 19980415 & 537.7 & 53\\
\textbf{51311} & 19990513 & 323.6 & 42 & \textbf{50980} & 19980616 & 410.5 & 49 & \textbf{50798} & 19971216 & 384.4 & 50\\
\textbf{51110} & 19981024 & 574.3 & 57 & \textbf{50960} & 19980527 & 391.2 & 50 & \textbf{50727} & 19971006 & 504.9 & 50\\
\end{tabular}
\end{table*}
\clearpage
\begin{table*}
\contcaption{ }
\begin{tabular}{rrrrrrrrrrrr}
Date & Date & Flux& Error&Date & Date & Flux& Error&Date & Date & Flux& Error\\
MJD  &  UT  & / mJy & / mJy&MJD  &  UT  & / mJy & / mJy&MJD  &  UT  & / mJy & / mJy\\\hline
\textbf{51114} & 19981028 & 2214.7 & 221 & \textbf{50800} & 19971218 & 2982.8 & 298 & \textbf{50641} & 19970712 & 2145.7 & 214\\
\textbf{51110} & 19981024 & 2588.8 & 258 & \textbf{50798} & 19971216 & 2658.6 & 265 & \textbf{50630} & 19970701 & 2082.6 & 208\\
\textbf{51102} & 19981016 & 2310.4 & 231 & \textbf{50793} & 19971211 & 2443.6 & 244 & \textbf{50623} & 19970624 & 2065.6 & 206\\
\textbf{51060} & 19980904 & 3428.8 & 342 & \textbf{50745} & 19971024 & 4642.5 & 464 & \textbf{50546} & 19970408 & 1878.4 & 187\\
\textbf{51059} & 19980903 & 3798.1 & 379 & \textbf{50744} & 19971023 & 4281.8 & 428 & \textbf{50545} & 19970407 & 1769.9 & 177\\
\textbf{51029} & 19980804 & 3861.0 & 386 & \textbf{50743} & 19971022 & 4247.9 & 424 & \textbf{50544} & 19970406 & 1741.3 & 174\\
\end{tabular}
\end{table*}
\clearpage
\begin{table*}
\contcaption{ }
\begin{tabular}{rrrrrrrrrrrr}
Date & Date & Flux& Error&Date & Date & Flux& Error&Date & Date & Flux& Error\\
MJD  &  UT  & / mJy & / mJy&MJD  &  UT  & / mJy & / mJy&MJD  &  UT  & / mJy & / mJy\\\hline
\hline\multicolumn{12}{c}{\textit{2318+049}}\\
\hline \textbf{51468} & 19991017 & 325.6 & 42 & \textbf{51010} & 19980716 & 363.9 & 47 & \textbf{50888} & 19980316 & 423.2 & 50\\
\textbf{51424} & 19990903 & 332.2 & 43 & \textbf{51009} & 19980715 & 329.8 & 42 & \textbf{50828} & 19980115 & 288.2 & 40\\
\textbf{51273} & 19990405 & 245.8 & 34 & \textbf{51006} & 19980712 & 367.1 & 47 & \textbf{50787} & 19971205 & 379.3 & 49\\
\textbf{51100} & 19981014 & 230.0 & 32 & \textbf{51004} & 19980710 & 342.0 & 44 & \textbf{50785} & 19971203 & 334.0 & 43\\
\textbf{51099} & 19981013 & 252.3 & 35 & \textbf{50999} & 19980705 & 388.6 & 50 & \textbf{50711} & 19970920 & 296.8 & 41\\
\textbf{51094} & 19981008 & 250.5 & 35 & \textbf{50980} & 19980616 & 414.4 & 49 & \textbf{51017} & 19980723 & 285.0 & 39\\
\hline\multicolumn{12}{c}{\textit{2345-167}}\\
\hline \textbf{51362} & 19990703 & 715.8 & 71 & \textbf{50727} & 19971006 & 893.2 & 89 & \textbf{50789} & 19971207 & 753.5 & 75\\
\end{tabular}
\end{table*}

\end{document}